\documentclass[aps,preprint]{revtex4}
\usepackage{graphicx}
\usepackage{amsmath}
\begin{document}
\title{\Large \bf  Note on Rotating Charged Black Holes in Einstein-Maxwell-Chern-Simons Theory}
\author{\large Alikram N. Aliev and Dilek K. \c Cift\c ci}
\address{Feza G\"ursey Institute, P. K. 6  \c Cengelk\" oy, 34684 Istanbul, Turkey}
\date{\today}

\begin{abstract}
We show that  the general solution of Chong, Cveti\u{c}, Lu and Pope for nonextremal rotating charged  black holes in five-dimensional minimal gauged supergravity, or equivalently in the Einstein-Maxwell-Chern-Simons theory with a negative cosmological constant and with the Chern-Simons coefficient $ \nu=1 $, admits a simple description in a  Kerr-Schild type framework with two scalar functions. Next, assuming this framework as an ansatz, we obtain new analytic solutions  for slowly rotating charged black holes in the Einstein-Maxwell-Chern-Simons theory with $ \nu\neq 1 .$ Using a  covariant superpotential derived from Noether identities within  the Katz-Bi\v{c}\'{a}k-Lynden-Bell  approach, we calculate  the mass  and angular momenta for the general supergravity  solution  as well as for the slowly rotating solution with two independent rotation parameters. For  the latter case, we also calculate the gyromagnetic ratios and  obtain  simple analytic formulas, involving both the parameters of the black holes and the Chern-Simons coefficient.
\end{abstract}

\maketitle

\section{Introduction}

The first study of higher-dimensional black hole solutions traces back to the work of Tangherlini \cite{tang}, who found an exact metric for static charged black holes, generalizing the Reissner-Nordstr\"om solution of four-dimensional Einstein-Maxwell theory to all higher dimensions. The Tangherlini  metric is also a solution to the Einstein-Maxwell-Chern-Simons theory  for any value of the Chern-Simons (CS) coefficient \cite{town}. However, the construction of rotating charged black hole solutions in the higher-dimensional Einstein-Maxwell theory turned out to be a rather complicated problem.  As is known, one of the most useful methods for constructing  such solutions in four dimensions  is based on the use of the Kerr-Schild form for the spacetime metric \cite{ks}. The remarkable property of the Kerr-Schild form is that the exact metric looks like its linearized approximation around the flat background spacetime. From the mathematical point of view, this property results in reduction of the Einstein-Maxwell equations to linear equations, thereby providing  a useful technique for solving these equations. However, in 1986 Myers and  Perry  showed that in higher dimensions this technique  works only in the uncharged case. They presented  the general asymptotically flat solution for rotating (non-charged) black holes in all spacetime dimensions \cite{mp}. Recently, the authors of work \cite{glpp} have shown that the previously-known Kerr-de Sitter metrics in four and five dimensions \cite{carter,hhtr} can also be put in the Kerr-Schild form, where  a ``linearization" occurs around de Sitter spacetime. Generalizing this framework to higher dimensions, they were able to present the general Kerr-de Sitter metrics in arbitrary spacetime dimensions.

A new  attempt to employ the Kerr-Schild framework  for constructing the rotating charged black hole solutions in higher-dimensional Einstein-Maxwell theory  was undertaken in \cite{aliev1, aliev2}. It was shown that this approach enables one to obtain  the higher-dimensional charged solutions only in the limit of slow rotation. Thus, the use of the Kerr-Schild framework to obtain the exact metrics for rotating black holes in  pure Einstein-Maxwell theory becomes restricted (at least, in its canonical form) to four spacetime dimensions. For this reason, the higher-dimensional counterpart of the Kerr-Newman solution still remains unknown.

Remarkably, the exact solutions for rotating charged black holes are known  in the Einstein-Maxwell theory with a Chern-Simons term. The addition to the theory of the Chern-Simons term with a particular value of the CS coefficient $ \nu=1 $, extends its symmetries, facilitating the search for the exact solutions. Furthermore,  for this particular value of the CS coefficient, the Einstein-Maxwell-CS theory with a negative cosmological constant in five-dimensions appears to be equivalent to five-dimensional minimal gauged supergravity. In the latter case, the rotating charged black hole solution with two equal angular momenta was constructed in \cite{clp}, while the general solution was found in \cite{cclp}. It is worth noting  that  these solutions were found in the absence of  solution-generating techniques. The authors have  employed only {\it a procedure of  trial and error} guided by their physical intuition. Later on, the rotating charged black hole  solutions with zero cosmological constant have also been obtained within generating techniques for solutions of five-dimensional minimal gauged supergravity \cite{galtsov, lp}.

The purpose of the present paper is two-fold: (i) to present a simple  Kerr-Schild type framework for rotating charged black holes in five-dimensional minimal gauged supergravity, (ii) to use this framework for constructing new analytic black hole solutions  to the Einstein-Maxwell-CS theory, when the CS coefficient $ \nu\neq 1 $.  It is known that  for  $ \nu=1 $ there exists a certain balance in distributions of the energy and angular momentum of these black holes, which is provided by supersymmetry \cite{town}. One can expect that going beyond  $ \nu=1 $ will violate this balance that may result in
new important features of these black holes, such as non-uniqueness  and instability. In this respect, the study of  the rotating charged black holes with $ \nu\neq 1 $ is of particular interest.

In Sec.II  we show that the spacetime of general rotating black holes in five-dimensional minimal gauged supergravity  admits two specific  congruences: a null  congruence and a spacelike  congruence. This allows one to  establish for these black holes the Kerr-Schild  type framework involving two scalar functions. In Sec.III we use the Kerr-Schild framework as an ansatz and obtain new analytic solutions  for slowly rotating charged  AdS  black holes of the Einstein-Maxwell-CS theory with $ \nu\neq 1 $. In Sec.IV employing the covariant  superpotential  technique  of Katz-Bi\v{c}\'{a}k-Lynden-Bell,  we calculate the mass and angular
momenta for the exact solution  with $ \nu=1 $ in the Kerr-Schild framework. Here we also calculate the mass, the angular momenta and  the gyromagnetic ratios  for the slowly rotating solution with two independent rotation parameters. In the Appendix  we give the components of the Einstein field equations calculated for the single rotation parameter case.

\begin{center}
\section{The Kerr-Schild framework of five-dimensional minimal supergravity}
\end{center}

The general metrics for rotating black holes in higher-dimensional gravity with a cosmological constant belong to the  Kerr-Schild class involving a single scalar function \cite{glpp}. It is also known that the general Kerr-Taub-NUT-de Sitter metrics in all higher dimensions admit a  multi-Kerr-Schild  structure \cite{ch}. For instance, the Kerr-Taub-NUT-de Sitter spacetime in five dimensions contains two linearly independent and mutually orthogonal null geodesic congruences. This enables one to put the spacetime  metric in a ``double" Kerr-Schild form  with two scalar functions \cite{cglp}. The extension of  the Kerr-Schild framework  to the case  of general rotating charged black holes in five-dimensional minimal gauged supergravity shows that the spacetime metric admits  two specific vector fields:  a null vector field $ k $ and a  spacelike vector field  $ \ell $, which allow us to present the  metric in the following form
\begin{equation}
ds^{2}=d\bar{s}^{2}+H k\otimes k + V\left(k \otimes \ell  + \ell\otimes k\right)\,, \label{ksframe1}
\end{equation}
where $ d\bar{s}^{2} $ is the  background AdS spacetime,  $ H  $ and  $ V  $ are two scalar functions. In the spacetime coordinates $ x^{\mu}=\{t, r, \theta, \varphi, \psi\}\,,~  \mu=0\,,...4 \,$,  the one-forms $ k $  and   $ \ell $ are given by
\begin{eqnarray}
k&=&k_{\mu}dx^{\mu}= \left\{\frac{\Delta _{\theta }}{\Xi _{a}\Xi _{b}}\, d t\,,\, 0\,,\, 0\,, \, -\frac{a\sin^{2}\theta}{
\Xi _{a}}\,d\varphi\,,\, -\frac{b\cos ^{2}\theta }{\Xi _{b}}\,d\psi \right\},\\ [4mm]
\ell&=& \ell_{\mu}dx^{\mu}=\left\{\frac{\Delta _{\theta }}{\Xi _{a}\Xi _{b}}\,\frac{a b}{l^2}\, dt\,,\, 0\,,\, 0\,, \, -\frac{b\sin^{2}\theta}{
\Xi _{a}}\,d\varphi\,,\, -\frac{a\cos ^{2}\theta }{\Xi _{b}}\,d\psi \right\}
\label{ln}
\end{eqnarray}
and the metric (\ref{ksframe1}) takes the form
\begin{eqnarray}
&&d{s}^{2} = \left[-\left(1+ \frac{r^{2}}{l^{2}}\right)\frac{\Delta _{\theta} }{\Xi_{a}\Xi _{b}}\,dt^{2}- 2 dr \left(\frac{ \Delta _{\theta} }{\Xi_{a}\Xi _{b}}\,dt -\frac{a\sin^{2}\theta}{
\Xi _{a}}\,d\varphi  -\frac{b\cos ^{2}\theta }{\Xi _{b}}\,d\psi \right) +\frac{\Sigma }{\Delta _{\theta }}\, d\theta ^{2}
\nonumber
\right. \\[4mm]  & & \left. \nonumber
+\,\frac{\left(r^{2}+a^{2}\right) \sin ^{2}\theta }{\Xi _{a}}\,d\varphi^2  +\frac{\left( r^{2}+b^{2}\right) \cos ^{2}\theta }{\Xi_{b}}\,d\psi^{2}\right] + H\left(\frac{\Delta _{\theta }}{\Xi _{a}\Xi _{b}}\, dt -\frac{a\sin^{2}\theta}{
\Xi _{a}}\,d\varphi  -\frac{b\cos ^{2}\theta }{\Xi _{b}}\,d\psi \right)^2 \\ [4mm] &&
+ 2 V\left(\frac{\Delta _{\theta }}{\Xi _{a}\Xi _{b}}\, dt -\frac{a\sin^{2}\theta}{
\Xi _{a}}\,d\varphi  -\frac{b\cos ^{2}\theta }{\Xi _{b}}\,d\psi \right)\left(\frac{\Delta _{\theta }}{\Xi _{a}\Xi _{b}}\,\frac{a b}{l^2}\, dt -\frac{b \sin^{2}\theta}{
\Xi _{a}}\,d\varphi  -\frac{a \cos ^{2}\theta }{\Xi _{b}}\,d\psi \right),
\label{ksframe2}
\end{eqnarray}
where  $ a $ and $ b $  are two independent rotation parameters, the expression in square brackets represents the AdS  spacetime with a length scale parameter $ l $, which is  determined by the negative cosmological constant $ l^2= - 6/\Lambda $. The metric functions
\begin{eqnarray}
\Delta_\theta & = & 1 -\frac{a^2}{l^2} \,\cos^2\theta
-\frac{b^2}{l^2} \,\sin^2\theta \,,~~~~~
\Sigma  =  r^2+ a^2 \cos^2\theta + b^2 \sin^2\theta \,
\label{ksmetfunc}
\end{eqnarray}
and
\begin{eqnarray}
\Xi_a &=&1 - \frac{a^2}{l^2}\,\,,~~~~~~~ \Xi_b=1 - \frac{b^2}{l^2}\,\,.
\label{xis}
\end{eqnarray}
It is straightforward to show that the vectors $ k_{\mu}$ and $ \ell _{\mu}$ satisfy the relations
\begin{eqnarray}
k_{\mu }k^{\mu }&=& 0 \,,~~~k_{\mu } \ell^{\mu }=0\,,~~~  \ell_{\mu } \ell^{\mu }= \frac{a^{2}\cos ^{2}\theta +b^{2}\sin ^{2}\theta }{r^{2}}
\label{norms}
\end{eqnarray}
with respect to both the background AdS spacetime and the full metric  (\ref{ksframe2}).  It follows that these two congruences are mutually orthogonal, but only one of them, namely the congruence  of the vector field $ k $ is null. This is in contrast to the ``double" Kerr-Schild metric form  of the Kerr-Taub-NUT-de Sitter spacetime  \cite{cglp}, where both geodesic vectors are null.

The potential one-form for the electromagnetic field of the spacetime  (\ref{ksframe1}) can be written  in terms of the null one-form $ k $. We have
\begin{equation}
A= \frac{\alpha}{\Sigma}\, \,k\,,
\label{potf}
\end{equation}
where $ \alpha $ is an arbitrary constant. The electromagnetic field two-form $ F=dA $ is given by
\begin{eqnarray}
F&=&\frac{2\alpha r}{\Sigma^2}\left(\frac{\Delta_\theta}{\Xi_a \Xi_b}\,dt-\frac{a \sin ^{2}\theta}{\Xi_a}\,d\varphi-\frac{b \cos ^{2}\theta}{\Xi_b}\,d\psi\right)\wedge dr\nonumber \\[3mm] && - \frac{\alpha\sin{2\theta}}{\Sigma^2}\left[ \frac{ (a^2-b^2)(1+ r^2/l^{2}) }{ \Xi_a \Xi_b}\,dt  - \frac{a (r^2+a^2)  }{ \Xi_a}\,d\varphi  + \frac{b(r^2+b^2)}{ \Xi_b}\,d\psi \right]\wedge d\theta\,.
\label{2form}
\end{eqnarray}
In constructing the Kerr-Schild framework  given in (\ref{ksframe1}),
we have used the known general solution for rotating charged black holes in five-dimensional minimal gauged supergravity \cite{cclp}. Indeed, with
\begin{eqnarray}
H &= & \frac{2M}{\Sigma}-\frac{Q^2}{\Sigma^2}\,\,,~~~~~ V= \frac{Q}{\Sigma}\,\,,~~~~~ \alpha=\frac{\sqrt{3}}{2}\,Q\,,
\label{hv}
\end{eqnarray}
where $ M $ is the mass and $ Q $  is the electric charge parameters
of the black hole, and  by applying to the metric (\ref{ksframe2}) the coordinate transformations
\begin{eqnarray}
dt&=&d\tau -\frac{(r^2+a^2)(r^2+b^2)+a b Q }{\Delta_r r^2}\,dr\,,\nonumber \\[2mm]
d\varphi&=&d\phi+  \frac{a}{l^{2}}\, d \tau-\frac{a(r^2+b^2)(1+ r^2/l^{2})+ Q b}{\Delta_r r^2}\,dr\,,\nonumber \\[2mm]
d\psi&=&d\chi+ \frac{ b}{l^{2}}\, d \tau-\frac{b(r^2+a^2)(1+ r^2/l^{2})+  Q a}{\Delta_r r^2}\,dr\,,
\label{transforms}
\end{eqnarray}
where
\begin{eqnarray}
\Delta_r &= &\frac{\left(r^2 + a^2\right)\left(r^2 +
b^2\right)\left(1+r^2/l^{2} \right)+ 2 a b Q + Q^2}{r^2} - 2 M \,, \label{delta}
\end{eqnarray}
we obtain the general solution of \cite{cclp} in rotating at infinity Boyer-Lindquist coordinates (see \cite{aliev3}). It is also important to note that with the metric (\ref{ksframe2}) and the potential one-form (\ref{potf})  one can easily solve the field equations of the Einstein-Maxwell-CS theory

\begin{equation}
R_{\mu}^{\,\nu}- 2\left(F_{\mu \lambda} F^{\nu
\lambda}-\frac{1}{6}\,{\delta_{\mu}}^{\nu}\,F_{\alpha \beta}
F^{\alpha \beta}\right) + \frac{4}{l^2}\,\delta_{\mu}^{\,\nu}=0\,,
\label{ein}
\end{equation}
\begin{equation}
\nabla_{\nu}F^{\mu\nu}+\frac{\nu}{2\sqrt{3}\sqrt{-g}}\,\epsilon^{\mu \alpha\beta\rho\tau} F_{\alpha\beta}F_{\rho \tau}=0\,,
\label{maxcs}
\end{equation}
for the supergravity value $ \nu=1 $, rederiving the solutions given in (\ref{hv}).

In the following,  we shall use the metric form  (\ref{ksframe2}) as an ansatz for constructing the rotating charged AdS black hole solutions with the CS coefficient  $ \nu\neq 1 $.

\vspace{4mm}

\section{Einstein-Maxwell-CS Black Holes}

As we have mentioned above, the values of the CS coefficient
make no difference to static Tangherlini type black holes of the Einstein-Maxwell-CS theory. However, this is not the case for stationary solutions. For instance, the value $ \nu= 1 $ turns out to be crucial for rotating  charged black holes described by the metric (\ref{ksframe2}). As far as we are aware, beyond this value of the CS coefficient, the exact rotating black hole solutions in five dimensions are not known. The authors of works \cite{kunz1,kunz2} have constructed  numerical solutions for the Einstein-Maxwell-CS  black holes, focusing on the case of spherical topology and two equal angular momenta. The numerical analysis uncovered  the new  interesting features of these black  holes, such as the lack of uniqueness property and the existence of unstable solutions.

In this section, restricting ourselves to the regime of slow rotation,
we give an analytic description of the  Einstein-Maxwell-CS  black holes with the  CS coefficient $ \nu\neq 1 $. For this purpose, we take the metric (\ref{ksframe2}) as a general ansatz  for putative new solutions and begin with calculating the contravariant components of the electromagnetic field tensor. For convenience, we take the same value for the parameter $ \alpha $ as given in (\ref{hv}). For the nonvanishing components, we  obtain
\begin{eqnarray}
F^{01}&=&-\frac{\sqrt{3}\, Q \left[(r^2+a^2)(r^2+b^2)+a b V \Sigma \right]} {r \Sigma^3}\,\,,
~~~~~~~~~~~ F^{02}=\frac{\sqrt{3}\, Q  (a^2-b^2) \sin2 \theta} {2 \Sigma^3}\,\,, \nonumber \\[3mm]
F^{13}&=&\frac{\sqrt{3}\, Q  \left[ a (r^2+b^2)(1+ r^2/l^{2})+b V \Sigma \right]}{r \Sigma^3}\,\,,~~~~~~~~~~~
F^{23}=-\frac{\sqrt{3}\, Q  a \Delta_\theta \cot \theta }{\Sigma^3}\,\,,
\nonumber \\[3mm]
F^{14}&=&\frac{\sqrt{3}\, Q  \left[b (r^2+a^2)(1+ r^2/l^{2})+a V \Sigma \right]}{r \Sigma^3}\,\,,~~~~~~~~~~
F^{24}=\frac{\sqrt{3}\, Q  b \Delta_\theta  \tan \theta}{\Sigma^3}\,\,.
\label{5demtcontra}
\end{eqnarray}
Substituting these components along with those given by (\ref{2form}) in the Maxwell-CS field equations (\ref{maxcs}), we find that they reduce to a single differential equation for the function $ V $. Thus, we have the equation
\begin{equation}
\frac{\partial V}{\partial r } - \frac{2 r}{\Sigma}\,V
 + \frac{4 Q \nu\, r}{\Sigma^2}  = 0\,,
\label{maxcs1}
\end{equation}
which admits a simple solution of the form
\begin{equation}
V=\frac{Q \nu}{\Sigma}\,.
\label{solution1}
\end{equation}
We note that this solution in general involves  an arbitrary function of $ \theta $ as well. However, the latter is fixed to be zero from the Einstein equations (\ref{ein}). In the following, to make the description more transparent, we  consider the solutions with single, two and equal rotation parameters  separately.

\subsection{The single rotation parameter solution with $ \nu\neq 1 $
}

Taking $ b=0 $ in the metric (\ref{ksframe2}) as well as in the expressions (\ref{2form}) and (\ref{5demtcontra}) for the electromagnetic field tensor, we calculate all the components of the field equations in (\ref{ein}) by  using Mathematica algebraic computing programme. Then, solving the  $ (X)^4_4 $ equation, we find the solution
\begin{equation}
H=\frac{2 M}{\Sigma} -\frac{Q^2\left(r^2 + a^2 \nu^2 \cos^2\theta\right)}{\Sigma^3}\,\,.
\label{solution44}
\end{equation}
Substituting  this solution in the remaining equations, we calculate explicitly their left-hand sides. The results are listed in the Appendix. From these expressions, it follows that  for the supergravity value $ \nu=1 $, the  function $ H $  in (\ref{solution44}) solves the complete set of equations in (\ref{ein}). Furthermore, these expressions  show that,  at the linear level in the rotation parameter $ a $ , the function $ H $  satisfies  the field equations for the value $ \nu\neq 1 $ as well. Thus, for $ \nu\neq  1 $, to within terms of linear order in $ a$, we have the solutions
\begin{eqnarray}
H &= & \frac{2M}{r^2}-\frac{Q^2}{r^4}\,\,,~~~~~ V= \frac{Q\nu}{r^2}\,.
\label{hvslow1}
\end{eqnarray}
Using these expressions in the metric (\ref{ksframe2}) and   passing to the Boyer-Lindquist coordinates in the linear in $ a $ approximation
\begin{eqnarray}
dt&=&d\tau -\frac{r^2}{\Delta}\,dr\,,\nonumber \\[2mm]
d\varphi&=&d\phi+  \frac{a}{l^{2}}\, d \tau -\frac{a (1+ r^2/l^{2})}{\Delta}\,dr\,,\nonumber \\[2mm]
d\psi&=&d\chi-\frac{Q a \nu }{\Delta r^2}\,dr\,,
\label{transforms1}
\end{eqnarray}
where
\begin{eqnarray}
\Delta &= & r^2 \left(1+r^2/l^2 \right)+ Q^2/r^2 - 2 M \,, \label{mdelta}
\end{eqnarray}
we put the metric in the form
\begin{eqnarray}
d{s}^{2} &=& -\frac{\Delta}{r^2}\, d\tau^2+ \frac{r^2}{\Delta}\,dr^2- 2 a \left(1-\frac{\Delta}{r^2}\right)\sin^2\theta d\phi\, d\tau
-\frac{2 Q a \nu}{r^2} \cos^2\theta d\chi\, d\tau + r^2 d\Omega^2_3\,\,,
\label{blframe1}
\end{eqnarray}
where
\begin{eqnarray}
d\Omega^2_3 &= & d\theta^2 + \sin^2\theta \,d\phi^2 + \cos^2\theta \,d\chi^2\,
\label{3sphere}
\end{eqnarray}
and the potential one-form of the electromagnetic field is given by
\begin{equation}
A = \frac{\sqrt{3} Q}{2 r^2}\, \left(d\tau - a \sin^2\theta d \phi \right)\,. \label{potform1}
\end{equation}
This metric describes slowly rotating charged Einstein-Maxwell-CS black holes with any value of the CS coefficient $ \nu\neq  1 $ within the linear in $ a $ approximation. For $ \nu=0  $ and for zero cosmological constant, it recovers the metric obtained earlier in \cite{aliev1, aliev2}  for  rotating charged black holes in pure Einstein-Maxwell theory. (See also a recent paper \cite{rong}).

\subsection{The two rotation parameters solution with $ \nu\neq 1 $ }

The generalization of the solution (\ref{blframe1}) to include two  independent  rotation parameters  is  straightforward. Indeed, checking the field equations (\ref{ein}) with the solutions (\ref{hvslow1}) in the regime of slow rotation we find that they are satisfied. Thus,  at the linear level in $ a $ and $ b $, the metric (\ref{ksframe2})  with the functions $ H $ and $ V $  given in  (\ref{hvslow1}) is also a solution to the Einstein-Maxwell-CS theory  with $ \nu\neq 1 $. Again, passing to the Boyer-Lindquist coordinates
\begin{eqnarray}
dt&=&d\tau -\frac{r^2}{\Delta}\,dr\,,\nonumber \\[2mm]
d\varphi&=&d\phi+  \frac{a}{l^2}\, d \tau-\frac{a r^2\left(1+ r^2/l^2\right)+ Q b \nu}{\Delta r^2}\,dr\,,\nonumber \\[2mm]
d\psi&=&d\chi+ \frac{b}{l^2}\, d \tau-\frac{b r^2\left(1+ r^2/l^2 \right)+ Q a \nu }{\Delta r^2}\,dr\,,
\label{transforms2}
\end{eqnarray}
we obtain  the metric in the form
\begin{eqnarray}
d{s}^{2} &=& -\frac{\Delta}{r^2}\, d\tau^2+ \frac{r^2}{\Delta}\,dr^2- 2 \left(1-\frac{\Delta}{r^2}\right)\left(a \sin^2\theta \,d\phi
+ b \cos^2\theta \,d\chi\right)d\tau \\ [4mm] &&
-\frac{2 Q \nu}{r^2} \left(b \sin^2\theta d\phi+ a \cos^2\theta d\chi\right)d\tau + r^2 d\Omega^2_3\,\,
\label{blframe2}
\end{eqnarray}
in which, the electromagnetic field is described by the potential one-form
\begin{equation}
A = \frac{\sqrt{3} Q}{2 r^2}\, \left(d\tau - a \sin^2\theta \,d \phi -
b \cos^2\theta\,d\chi\right)\,. \label{potform2}
\end{equation}
For $ \nu=0 $ and $ l\rightarrow \infty $, this metric agrees with that obtained in \cite{aliev1}.

\subsection{The solution with equal rotation parameters}

We now discuss an example of another class of solutions  with $ \nu\neq 1 $, which, unlike the solutions given above, exhibits the singularity in the regime of slow rotation. Following the same strategy as in the previous subsections, we first write down  explicitly the complete set of the Einstein field equations (\ref{ein}), using
the metric ansatz  (\ref{ksframe2})  and the electromagnetic field components given in (\ref{2form}) and (\ref{5demtcontra}) along with (\ref{solution1}). Next, solving  the  $ (X)^2_2 $  equation, we find
\begin{eqnarray}
H=\frac{2 M}{\Sigma} + \frac{Q^2}{\Sigma^2}\,\left(1-2 \nu^2\right)\left[1+ \frac{2(1-\nu^2)}{1-2 \nu^2}\frac{\Sigma}{\Sigma-r^2} \ln \left(\frac{r^2}{2\Sigma}\right)\right] .
\label{solution22}
\end{eqnarray}
Substituting this solution in the remaining equations, we obtain that they are satisfied only for slow rotation with equal rotation parameters. Thus, the solution subject to the  system of the Einstein-Maxwell-CS  equations  with  $ \nu\neq 1 $ is given by
\begin{eqnarray}
H=\frac{2 M}{r^2} - \frac{Q^2}{r^4}\left[1- 2\ln2 \, (1-\nu^2)\left(1-\frac{r^2}{a^2} \right)\right]\,,
\label{slow22}
\end{eqnarray}
which is obtained from (\ref{solution22}) by taking  $ a= b $  and expanding it in powers of the rotation parameter up to the linear order. We note that  the expression for the function $ V $  is the same as that given in  (\ref{hvslow1}).  Using the coordinate transformations
\begin{eqnarray}
dt&=&d\tau -\frac{dr}{f(r)}\,\,,\nonumber \\[2mm]
d\varphi&=&d\phi+  \frac{a}{l^2}\, d \tau-\frac{a}{r^2}\,\frac{dr}{g(r)}\,,\nonumber \\[2mm]
d\psi&=&d\chi+ \frac{a}{l^2}\, d \tau -\frac{a}{r^2}\,\frac{dr}{g(r)}\,\,,
\label{transforms3}
\end{eqnarray}
where
\begin{eqnarray}
f(r)&= & 1+\frac{r^2}{l^2} - H\,,~~~~~~~ g(r)= f(r)\left(1+\frac{r^2}{l^2} + V\right)^{-1},
\label{fg}
\end{eqnarray}
we obtain that the spacetime metric in the Boyer-Lindquist coordinates takes the form
\begin{eqnarray}
d{s}^{2} &=& - f(r) d\tau^2 + \frac{dr^2}{f(r)}
- 2 a \left[1-f(r) +V \right]\left(\sin^2\theta d\phi+ \cos^2\theta d\chi\right)d\tau + r^2 d\Omega^2_3\,,
\label{blframe3}
\end{eqnarray}
where $ H $ is given by (\ref{slow22}). The potential one-form is  given by (\ref{potform2}) with $ a= b $. We see that for $ a\rightarrow 0 $ the metric components,  unlike those in  (\ref{blframe2}), become divergent. That is, the metric exhibits  the singularity (instability)  with respect to slow rotation. We conclude that the analytical examples of metrics given above explicitly show  that the physics of rotating black holes in the Einstein-Maxwell-CS theory crucially depends on the value of the CS coefficient. These results are in  qualitative agreement with the numerical  analysis  of works \cite{kunz1,kunz2}.

\begin{center}
\section{The masses, angular momenta and the gyromagnetic ratios}
\end{center}

The most simple  way of calculation the mass and angular momenta of black holes  in asymptotically flat spacetime is achieved within the Komar approach \cite{komar}. However, this approach must be used with care for rotating AdS black holes, where  it gives an ambiguous result for the mass. Therefore, the mass of these black holes  has been calculated by integrating the first law of thermodynamics \cite{gpp1} as well as in the framework of other approaches based on the use of conservation laws derived from the symmetries of the system \cite{katzder, dkt, aliev4}.  The mass of rotating charged black holes in five-dimensional minimal gauged supergravity has also been calculated using both the first law of thermodynamics \cite{cclp} and the conformal definition of Ashtekar, Magnon  and Das \cite{chen}. Here we wish to calculate the mass of these black holes using the Kerr-Schild framework  (\ref{ksframe2}) for the spacetime metric and employing the covariant superpotential technique of Katz, Bi\v c\' ak and Lynden-Bell (KBL) \cite{kbl}.(See also a recent review \cite{petrov1}).  Using the language of differential forms, we can write the integral of the KBL superpotential in the form
\begin{equation}
K=- \frac{1}{16 \pi}\oint\, ^{\star}d(\delta \hat \xi) - \frac{1}{8
\pi} \oint \,^{\star}(\delta S) \,\,, \label{kblpot}
\end{equation}
where
\begin{eqnarray}
S&= &\frac{1}{2} \,\xi_{[\mu} \kappa_{\nu]}\, d x^{\mu}\wedge
dx^{\nu}\,\,,~~~~~~~~~\kappa^{\mu}= g^{\mu \nu }\, \delta
\Gamma^{\lambda}_{\nu \lambda}- g^{\alpha \beta }\, \delta
\Gamma^{\mu}_{\alpha \beta }\,\,
\end{eqnarray}
and $\,\delta \hat \xi \,$  denotes the difference between the Killing isometries of the original spacetime (\ref{ksframe2}) and its reference background. The latter is  obtained  by taking $ M=0 $ and $ Q=0 $ in the original spacetime. Similarly,  $\,\delta\Gamma^{\lambda}_{\sigma \rho}\,$ stands for the difference between the Christoffel symbols of the original spacetime and those of its reference background. The difference between the corresponding metric determinants $\,\delta g=0\,$.

Evaluating the integral in (\ref{kblpot}) with respect to the timelike Killing vector $ \partial_t $  and over a $3 $-sphere at spatial
infinity, we obtain the mass $ M^{\prime}= K [\partial_t] $ of the spacetime  (\ref{ksframe2}). Indeed, using  for  the integrands  the asymptotic expansions at $ r \rightarrow \infty $
\begin{eqnarray}
&&\delta \xi^{t\,;\,r}_{(t)}=\frac{2}{\Xi_a  \Xi_b}\,\left[\frac{M\left(2\, \Xi_b \sin^2{\theta}+2\, \Xi_a \cos^2{\theta}-\Xi_a \Xi_b\right)+ 2 Q a  b \, l^{-2} \Delta_{\theta}}{r^3}\right]+\mathcal{O}\left(\frac{1}{r^5}\right)\,\,,
\\[4mm]
&&\xi^{[t}\kappa^{r]}=-\frac{M}{r^3}+\mathcal{O}\left(\frac{1}{r^5}\right)\,\,, \label{expansions1}
\end{eqnarray}
and performing the integration procedure, we find  the expression
\begin{equation}
M^{\prime}=  \frac{\pi M \left(2 \Xi_a +2 \Xi_b-
\Xi_a\ \Xi_b\right) + 2 \pi Q a b \, l^{-2}\left(\Xi_a+\Xi_b\right)}{4\,\Xi_{a}^2 \Xi_{b}^2}
\,\,,\label{mass1}
\end{equation}
which is precisely the same  as  that obtained earlier within other  approaches \cite{cclp, chen}.

The angular momenta of the spacetime  (\ref{ksframe2}) are obtained by evaluating the integral in (\ref{kblpot}) with respect the rotational Killing vectors $\partial_ \phi\,$ and $ \partial_ \psi\,$. Namely, we have $ J^{\prime}_{a}=- K\left[\partial_{\phi}\right] $  and $ J^{\prime}_{b}=- K\left[\partial_{\psi}\right] $. Substituting the asymptotic expansions at $ r \rightarrow \infty $
\begin{eqnarray}
&&\delta \xi^{t\,;\,r}_{(\varphi)}=- \frac{2 \sin^2\theta}{\Xi_a}\,\, \left[ \frac{2 a M + Q b (2- \Xi_a)}{r^3} \right] +\mathcal{O}\left(\frac{1}{r^5}\right)\,,\\[3mm]
&&\delta \xi^{t\,;\,r}_{(\psi)}=- \frac{2 \cos^2\theta}{\Xi_b}\,\, \left[ \frac{2 b M + Q a (2- \Xi_b)}{r^3} \right] +\mathcal{O}\left(\frac{1}{r^5}\right)\,,
\label{expansions2}
\end{eqnarray}
in  (\ref{kblpot})  and taking the integrals over a $3 $-sphere at spatial infinity, we obtain that the angular momenta are given by
\begin{eqnarray}
J^{\prime}_{a} &= &\frac{\pi}{4}\,\, \frac{2 a M+ Q b (2-\Xi_a)}{\Xi_{a}^2
\Xi_b}\,,\\[3mm]
J^{\prime}_{b} &= &\frac{\pi}{4}\,\,\frac{2 b M + Q a (2-\Xi_b)}{\Xi_{a}
\Xi_{b}^2}\,.
\label{angjj1}
\end{eqnarray}
These expressions also agree with those obtained in \cite{cclp, chen}.

Next, we calculate the mass and angular momenta for the  spacetime (\ref{blframe2}) of  slowly rotating charged black holes in the Einstein-Maxwell-CS theory with  $ \nu\neq 1 $. Performing for this case similar calculations  of the integral in (\ref{kblpot}), we find
\begin{eqnarray}
M^{\prime} &= &\frac{3\pi}{4} M\,,~~~~~~~J_a=\frac{\pi}{4} \left(2a M + Q b \nu\right)\,,~~~~~~~ J_b=\frac{\pi}{4} \left(2 b M + Q a \nu\right)\,.
\label{msjj2}
\end{eqnarray}
It follows that the mass of these  black holes is not affected by the CS coefficient, whereas their angular momenta depend on the CS coefficient. It is also interesting  to know how the CS coefficient affects the gyromagnetic ratios of  these black holes. For  $ \nu=1 $,
the gyromagnetic ratios for  the general metric (\ref{ksframe2}) was calculated in \cite{aliev3}. Using the  arguments of this work for the metric (\ref{blframe2}), we can  define   the two gyromagnetic ratios
\begin{eqnarray}
g_{a} & = & \frac{2 M^{\prime} \mu^{\prime}_{a}}{Q^{\prime} J^{\prime}_{a}} \,\,,~~~~~~~~g_{b}  = \frac{2 M^{\prime} \mu^{\prime}_{b}}{Q^{\prime} J^{\prime}_{b}}\,\,,
\label{twogyros}
\end{eqnarray}
where
\begin{eqnarray}
\mu^{\prime}_{a} & = & Q^{\prime} a \,,~~~~~~~~~~ \mu^{\prime}_{b}  =  Q^{\prime} b\,\,
\label{dipoles}
\end{eqnarray}
are two magnetic dipole moments  of the black hole associated with its two rotation $2$-planes and $  Q^{\prime} $ is the physical charge  determined by the Gauss law. Substituting in (\ref{twogyros}) the expressions for the mass and angular momenta from  (\ref{msjj2}), we  find
\begin{eqnarray}
\label{zerolambda1}
g_{a} & = &  3 \left(1- \frac{Q b\nu}{2aM + Q b\nu}\right)\,,
\\  [2mm]
g_{b} & =&   3 \left(1- \frac{Q a\nu}{2bM + Q a\nu}\right)\,.
\label{zerolambda2}
\end{eqnarray}
It is easy to see that for $ \nu \rightarrow 0 $, the gyromagnetic ratio tends to its value \cite{ aliev1} in the Einstein-Maxwell theory, $ g  \rightarrow  3 $ . From these expressions, it also follows  that
for given parameters of the black hole, the value of the gyromagnetic ratios tend to decrease with the growth (within the linear approximation in rotation parameters) of the CS coefficient.

\vspace{4mm}

\section{Conclusion}

The Kerr-Schild framework played a profound role in classical  general relativity, resulting in the discovery of the most important stationary black hole solution to the Einstein field equations, which nowadays is known as the Kerr metric.  It turned out that the Kerr-Schild framework survives in higher-dimensional general relativity as well, thereby paving the way for the construction of the counterpart of the Kerr and Kerr-de Sitter metrics in all higher dimensions.

In this paper, we have shown that the Kerr-Schild framework can also be extended to five-dimensional minimal gauged supergravity, where  the
general metric for rotating charged black holes admits  a  Kerr-Schild type form with two scalar functions. It should be stressed that in this metric form only one of two congruences is null, the other one is spacelike. This framework ensures the simple derivation of the general rotating charged black hole solution in five-dimensional minimal gauged supergravity. Assuming the ``double"  Kerr-Schild form as an ansatz, we have found new analytic solutions  for slowly rotating charged AdS black holes of the Einstein-Maxwell-CS theory with $ \nu\neq 1 $. These are, to our knowledge, the first examples of analytic solutions in the rotating case.

Using the covariant  superpotential  technique  of Katz-Bi\v{c}\'{a}k-Lynden-Bell, we have  calculated the mass and angular momenta for both  the general black holes with $ \nu=1 $ and the slowly rotating black hole with any value of the CS coefficient $ \nu\neq  1 $ within the linear approximation in slow rotation. We have also given simple analytic expressions for the gyromagnetic ratios of the slowly rotating Einstein-Maxwell-CS black holes.

\section{Acknowledgments}
A. N. thanks  Teoman Turgut and Nihat Berker for their  stimulating encouragements.  D. K.  thanks T{\"U}B\.{I}TAK (BIDEB-2218)  for financial support.

\appendix*

\section{The Field Equations}
Introducing for convenience the notation
\begin{equation}
M_{\mu}^{\,\nu}=F_{\mu \lambda} F^{\nu
\lambda}-\frac{1}{6}\,{\delta_{\mu}}^{\nu}\,F_{\alpha \beta}
F^{\alpha \beta}
\label{mm}
\end{equation}
in the field equations (\ref{ein}) and using the solution of the $ (X)^4_4 $ equation given by (\ref{solution44}), we obtain
 \begin{displaymath}
R^{0}_{0}- 2 M^{0}_{0}+\frac{4}{l^2}=- \frac{4 Q^2 a^2 (\nu^2-1)\cos^2\theta}{\Xi_a \,\Sigma^6}\left[ 4 r^2 \Delta_{\theta}(r^2+a^2) +  \Sigma \,a^2 \sin^2\theta \left(1+\frac{2r^2}{l^2}\right)\right],
\label{eq00}
\end{displaymath}
\begin{displaymath}
R^{1}_{1}- 2 M^{1}_{1}+\frac{4}{l^2}=- \frac{16 Q^2 a^2 r^2 (\nu^2-1)\cos^2\theta}{\Sigma^5}\,,~~~~R^{2}_{2}- 2 M^{2}_{2}+\frac{4}{l^2}  =   \frac{4 Q^2 a^2 (\nu^2-1)\cos^2\theta}{\Sigma^5}\,(r^2+\Sigma)\,,
\label{eq11}
\end{displaymath}
\begin{displaymath}
R^{3}_{3}- 2 M^{3}_{3}+\frac{4}{l^2}=\frac{4 Q^2 a^2 (\nu^2-1)\cos^2\theta}{\Xi_a \Sigma^6}
\left\{\Delta_{\theta} \left[\Sigma a^2 + 2 r^2 (r^2+a^2)\right]
+ 2 a^2 r^2 \sin^2\theta \left(1+\frac{r^2}{l^2}+\frac{\Sigma}{l^2}
\right)\right\},\\[4mm]
\label{eq33}
\end{displaymath}
\begin{displaymath}
R^{1}_{3}= -\frac{4 Q^2 a^3 r^2(\nu^2-1)\sin^2\theta}{\Xi_a \,\Sigma^6}\left[ \Delta_{\theta}(r^2 \cos2\theta + a^2 \cos^2\theta) +\Xi_a \Sigma  \cos^2\theta+\frac{a^2 \sin^2 2\theta }{2}\,\left(1+\frac{r^2}{l^2}\right)\right],
\label{eq13}
\end{displaymath}
\begin{eqnarray}
R^{2}_{3}&=&\frac{ Q^2 a^3 r (\nu^2-1) \left(3\Sigma- 8 r^2\right)\Delta_{\theta} \sin^2\theta \sin 2\theta}{\Xi_a \Sigma^6} \,,~~~~~~~R^{2}_{0}=- R^{2}_{3}\,\frac{\Delta_{\theta}}{a \sin^2\theta} = - R^{1}_{2}\,\frac{\Delta^2_{\theta}}{\Xi_a \Sigma}\,\,,\nonumber\\[6mm]
R^{1}_{0}&= &\frac{4 Q^2 a^2 r^2(\nu^2-1)\Delta_{\theta}}{\Xi_a \Sigma^6}\left[ \Delta_{\theta}(r^2 \cos2\theta + a^2 \cos^2\theta) +  \frac{a^2 \sin^2 2\theta }{2}\,\left(1+\frac{r^2}{l^2}\right)\right]\nonumber\,,
\label{eq23}
\end{eqnarray}
\begin{displaymath}
R^{4}_{3}- 2 M^{4}_{3}=\frac{ 4 Q^3 a^4 \nu (\nu^2-1)\sin^2 2\theta}{\Xi_a \Sigma^6}\,\,,~~~~~~~~~~R^{4}_{0}- 2 M^{4}_{0}=-\left(R^{4}_{3}- 2 M^{4}_{3}\right) \frac{\Delta_{\theta}}{a \sin^2\theta}\,\,,
\label{eq43}
\end{displaymath}
\begin{eqnarray}
R^{3}_{0}- 2 M^{3}_{0}&=&-\frac{ 4 Q^2 a^3 (\nu^2-1)\Delta_{\theta}\cos^2\theta}{\Xi_a \Sigma^5}\left[\frac{r^2}{l^2}+ \left(1+\frac{r^2}{l^2}\right)\left(1+\frac{4r^2}{\Sigma}\right)\right],\nonumber\\[4mm]
R^{0}_{3}- 2 M^{0}_{3}&=&\frac{ Q^2 a^3 (\nu^2-1)\sin^2 2\theta}{\Xi_a \,\Sigma^5}\left[ r^2 +(r^2+a^2) \left(1+\frac{4r^2}{\Sigma}\right)\right]\,.
\label{eq30}
\end{eqnarray}
We note that  the above expressions vanish identically for the values $\nu= \pm1 $.  Furthermore, we note that they involve only the square or higher powers of the rotation parameter $ a $. It follows that for any  $ \nu\neq  \pm1 $, to within the linear approximation in $ a $, these expressions vanish  as well. That is,   the solution in (\ref{solution44}) satisfies the complete set of the field equations (\ref{ein}) in the limit of slow rotation.

\end{document}